\documentstyle[aps,epsf]{revtex}
\begin{document}
\draft\twocolumn
\title{Comment on ``Loss-error compensation in quantum-state measurements''}
\author{G. M. D'Ariano$^{a}$ and C. Macchiavello$^{b}$}
\address{$^a$ Dipartimento di Fisica \lq Alessandro Volta\rq, Universit\`a 
degli 
\nobreak Studi di Pavia, via A. Bassi 6, I-27100 Pavia, Italy}
\address{$^b$ Clarendon Laboratory, University of Oxford, Parks Road, 
OX1 3PU, Oxford, UK}
\maketitle
\begin{abstract}
In the two papers [T. Kiss, U. Herzog, and U. Leonhardt, Phys. Rev. A
{\bf 52}, 2433 (1995); U. Herzog, Phys. Rev. A {\bf 53}, 1245 (1996)]
with titles similar to the one given above, the authors assert that in
some cases it is possible to compensate a quantum efficiency $\eta\leq
1/2$ in quantum-state measurements, violating the lower bound $1/2$
proved in a preceding paper [G. M. D'Ariano, U. Leonhardt and H. Paul,
Phys. Rev. A {\bf 52}, R1801 (1995)]. Here we re-establish the bound
for homodyning any quantum state, and show how
the proposed loss-compensation method would fail in a
real measurement outside the $\eta >1/2$ regime.
\end{abstract}
\bigskip
In Ref. \cite{DLP} the homodyne tomography method---so far the only
experimental scheme to measure the density matrix of the quantum state
of radiation---was  shown to be robust 
to the smearing effect of nonunit quantum efficiency $\eta$ at detectors. 
It was proved that the measurement is possible only above a lower bound for 
$\eta$ that
depends on the chosen matrix representation of the state $\hat\varrho$.
In particular, for the Fock basis, such bound is $\eta=1/2$, and 
in the same Ref. \cite{DLP} a reconstruction algorithm that depends
parametrically on $\eta$ was provided. 
As noticed in Ref. \cite{Yuen},
the existence of a lower bound for quantum efficiency in state
measurements is a relevant issue for the foundations of
quantum mechanics, as it prevents the measurement
of the wave function of an individual quantum system through a series
of repeated weak measurements. Therefore, if a scheme for
loss-error compensation is devised, then, in principle, such scheme can
open the door to prove an inconsistency within the logical framework of
quantum mechanics.
\par In Ref. \cite{KHL} it was proposed to perform the state
measurement by using the same algorithm of Ref. \cite{DLP} for 
$\eta=1$, and treating the effect of a low quantum efficiency as a
lossy evolution of radiation, thus separating the ``bare'' detection
with $\eta=1$ from the loss-compensation procedure. More explicitly,
the idea is to regard any state measurement with $\eta<1$ on the
``signal'' state  $\hat\varrho_{sig}$ as the corresponding
hypothetical bare measurement with $\eta=1$ on a ``dressed'' damped
state $\hat\varrho_{meas}$. Also, stated in different 
words, the effect of nonunit quantum efficiency is referred to the
quantum state itself rather than to the detector, regarding non unit
quantum efficiency in a Schr\H{o}dinger-like picture, with the state
evolving from $\hat\varrho_{sig}$ to $\hat\varrho_{meas}$, 
and the quantum efficiency playing the role of a time parameter $t=-\ln\eta$.
The core of the method is the inversion of the generalized
Bernoulli loss-transformation, which relates the matrix elements of the 
signal state $\hat\varrho_{sig}$ with those of the dressed state 
$\hat\varrho_{meas}$ in the Fock basis. This is given by
\begin{eqnarray}
&&\langle n|\hat\varrho_{sig}|n+d\rangle=
{{\eta^{-\frac{1}{2}(2n+d)}}\over{\sqrt{n!(n+d)!}}}
\sum_{j=0}^\infty {{\sqrt{(n+j)!(n+d+j)!}}\over{j!}} \nonumber\\
&&(1-\eta^{-1})^j \langle n+j|\hat\varrho_{meas}|n+d+j\rangle
\;.\label{Bernoulli}
\end{eqnarray}
The argument of Ref.\cite{KHL} states that above the bound $\eta=1/2$
the convergence of series (\ref{Bernoulli}) is guaranteed
for any state, but the convergence radius for $\eta$ depends on
the matrix elements of $\hat\varrho_{meas}$, and, in principle, for some
particular states it is possible for the series to converge also for
$\eta\le 1/2$. Thus, for example, for a thermal state with $\bar{n}$
photons, the matrix elements of $\hat\varrho_{meas}$ decay as
$[\bar{n}/(\bar{n}+1)]^j$ versus the summation index $j$, and one
concludes that 
the series (\ref{Bernoulli}) converges for $\eta>(2+1/\bar{n})^{-1}$,
which violates the bound $\eta=1/2$. 
In this comment we show that this argument doesn't
apply to a measurement, because in this case the series (\ref{Bernoulli}) 
must be evaluated with coefficients given by the measured values of
the matrix elements of $\hat\varrho_{meas}$ in place of their expectation 
values $\langle n|\hat\varrho_{meas}|m\rangle$. 
In order to have a successful experiment based 
on $N$ repeated measurements, the series with 
measured coefficients must converge for increasingly large truncation
index to a result within an error that vanishes for large $N$.   
For statistically uncorrelated coefficients $c_j$
the variance of the (unconditionally) convergent series 
$\sum_{j=0}^{\infty} z^jc_j$ is given by the series of variances
$\sum_{j=0}^{\infty} z^{2j}\langle\Delta c^2_j\rangle$,
where $z=1-\eta^{-1}$. Then, an {\em a
priori} estimation of the measurement error is given by 
$\epsilon=\sqrt{\sum_{j=0}^{\infty} z^{2j}\epsilon^2_j}$, with
$\epsilon_j=\sqrt{\langle\Delta c^2_j\rangle/N}$. 
Now, it is clear that statistical errors depend on the particular
detection scheme used to measure $c_j\propto\langle
n+j|\hat\varrho_{meas}|n+d+j\rangle$. For direct
photodetection---which, however, is not a quantum-state measurement---the 
error $\epsilon_j$ associated to a diagonal element $p_j=\langle
n+j|\hat\varrho_{meas}|n+j\rangle$ is $\epsilon_j\simeq\sqrt{(1-p_j)p_j/N}$, 
and $\epsilon_j^2$ vanishes linearly with $p_j$ itself for $j\to\infty$.
This is the logical basis of the argument of
the commented Refs. \cite{KHL,H}, \cite{nota1}, where,
however, the effects of statistical errors in homodyne tomography were not
considered. Actually, when the density matrix
elements are measured
 by homodyne tomography, the statistical error $\epsilon_j$
does not vanish as the respective matrix element $p_j=\langle
n+j|\hat\varrho_{meas}|n+d+j\rangle$ (for fixed $n$ and $d$), but 
`saturates' to the value $\sqrt{2/N}$, independently of $d$, $n$, and
$\hat\varrho_{meas}$ \cite{optcomm,nota}. 
Therefore, the above argument no longer
holds, and the convergence radius equals the lower bound $\eta=1/2$. 
Thus, for $\eta\leq 1/2$ the loss-compensation procedure is meaningless for any
state, even though the series (\ref{Bernoulli}) converges
analytically, as for the case of the thermal states considered before. 
The same is also true for the iterated analytical continuation procedure given
by Eq. (14) of Ref. \cite{H}. Here, one has a number of infinite
series that involve matrix elements of the bare state with shifted
indices. However, in an actual measurement, a truncation value for each
series must be given, and this is determined by the largest
index of the measured density matrix. 
With such truncation one can reorder terms
in the series and prove very easily that Eq. (14) of Ref. \cite{H} 
and Eq. (\ref{Bernoulli}) are exactly the same. For this reason, in
the following we will always refer only to the loss-error compensation
procedure (\ref{Bernoulli}) of Ref. \cite{KHL}. 
\begin{figure}[hbt]\vskip .3truecm\begin{center}
\epsfxsize=.65\hsize\leavevmode\epsffile{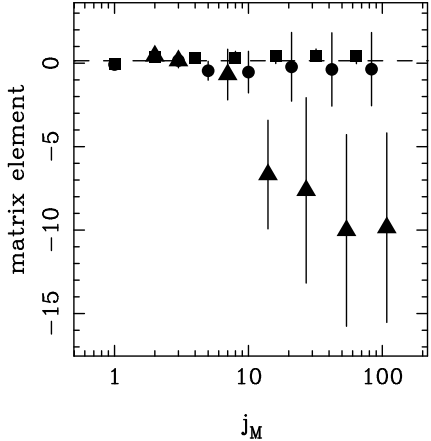}\end{center}
\caption{Matrix element $\langle 2|\hat\varrho_{sig}|2\rangle$
for a thermal state with 2 photons
evaluated as the truncated series (\protect\ref{Bernoulli}) 
with homodyne detected matrix elements $\langle
j+2|\varrho_{meas}|j+2\rangle$ versus the truncation 
index $j_M$. The squares correspond to
$\eta=.6$, the circles to $\eta=.55$, the triangles to $\eta=.53$. All
points are obtained with Monte Carlo experiments 
using 24000 homodyne data.}\label{f:fig1}\end{figure}

\par We illustrate the effect of experimental errors in the
loss-compensation procedure by means of Monte Carlo simulated
experiments of homodyne tomography with $\hat\varrho_{sig}$ as a thermal 
state with $\bar n=2$. According to Refs. \cite{KHL,H} this state should be
accessible for $\eta>2/5$, hence outside the allowed region $\eta>1/2$.

\begin{figure}[hbt]\vskip .3truecm\begin{center}
\epsfxsize=.65\hsize\leavevmode\epsffile{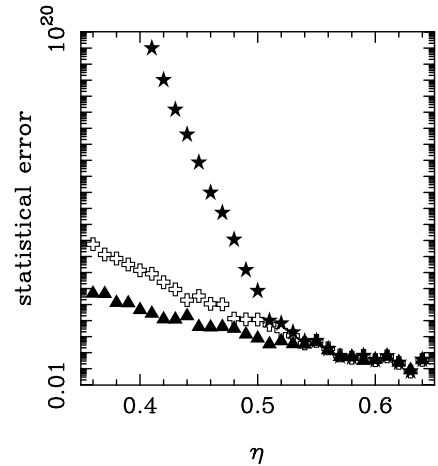}\end{center}
\caption{Statistical error corresponding to the same matrix element 
and the same state as in Fig. \protect\ref{f:fig1}, evaluated as the
truncated series (\protect\ref{Bernoulli}), as a function of $\eta$ for
different values of the truncation index $j_M$. The triangles correspond
to $j_M=10$, the crosses to $j_M=20$ and the stars to $j_M=100$.
All points are obtained with Monte Carlo experiments
using 8000 homodyne data.}
\label{f:fig2}\end{figure}

In Fig. \ref{f:fig1} we report the matrix element 
$\langle 2|\varrho_{sig}|2\rangle$ evaluated as the
truncated series (\ref{Bernoulli}) with homodyne detected matrix
elements $\langle j+2|\varrho_{meas}|j+2\rangle$ versus the truncation
index $j_M$ of the series, and for different values of $\eta$. One can
see that for $\eta$ well above the bound $\eta=1/2$ ($\eta=.6$ and
$\eta=.55$) the series converges versus $j_M$ to the correct
theoretical value, whereas for $\eta$ approaching the bound the size
of the error bar increases dramatically, with the sum departing more and
more from the theoretical value. For $\eta=1/2$ or below, the result
is out of the scale of the figure and oscillates unboundedly versus
$j_M$. In Fig. \ref{f:fig2} the statistical error for the same matrix element 
is studied versus the quantum efficiency $\eta$ for
different values of the truncation index $j_M$:
it is apparent that the error converges for all values $\eta>1/2$. 
When the bound $\eta=1/2$ is approached, the error starts
growing as a function of the truncation index $j_M$, and diverges
for all values $\eta\leq 1/2$ (the finite slope versus $\eta$ is only
due to the finiteness of $j_M$). 
As we can see from the figure, 
$\eta=1/2$ is manifestly a ``transition value'' from convergent to
divergent behavior of the statistical error.

Actually, in a real experiment homodyne data are contaminated by
additional sources of noise other than quantum efficiency. However,
for tomography some kinds of noise can still be treated as a negative
contribution to the overall quantum efficiency \cite{Tokyo}.

Before concluding, we want to stress that to our knowledge there is no
method---alternative to homodyne tomography---that has statistical
errors which make the series (\ref{Bernoulli}) convergent for $\eta<1/2$.
For all methods recently proposed by several authors
\cite{bar,banasek,paul}, an analysis of statistical errors is still 
lacking. Moreover, in the set of papers \cite{bar} the radiation is
measured indirectly through measurements on atoms that interacted with
it, and in such circumstance the quantum noise/loss of the apparatus
cannot be described just in terms of an overall quantum efficiency.

In conclusion, we have shown that the loss compensation procedures
proposed in Refs. \cite{KHL,H} cannot be used to measure the quantum
state through homodyne tomography below the bound $\eta=1/2$ proved in
Ref. \cite{DLP}. 
These procedures are still valid for $\eta<1/2$ 
when the diagonal matrix elements  are measured by direct detection.
 
C.M. acknowledges support by the European Union TMR Programme.
This work was supported in part by the TMR Research Network ERP-4061PL95-1412

\begin{references}
\bibitem{nota1} Notice, however, that the convergence radius $z<\sqrt R$
is the square root of the convergence radius of the series
(\protect\ref{Bernoulli}).
\end{references}
\end{document}